\documentclass[aps,prd,twocolumn,showpacs,superscriptaddress]{revtex4-1}

\usepackage{epsfig,amsmath}
\usepackage[english]{babel}
\usepackage{pgf,pgfarrows,pgfnodes,pgfautomata,pgfheaps}
\usepackage{amsfonts,amsmath,amsthm,amssymb,enumerate,array,amssymb}
\usepackage[latin1]{inputenc}
\usepackage{bm}
\usepackage{fixmath}
\usepackage{amsbsy}
\usepackage{graphicx}
\usepackage{capt-of}

\usepackage[normalem]{ulem}

\begin{document}

\title{Circular geodesics of naked singularities
in the Kehagias-Sfetsos metric of Ho{\v r}ava's gravity} 

\author{Ronaldo S. S. Vieira}\email[]{ronssv@ifi.unicamp.br}
\affiliation{Instituto de F{\'{\i}}sica ``Gleb Wataghin'',
  Universidade          Estadual de Campinas, 13083-859, Campinas, SP,
  Brazil}    
\affiliation{Copernicus Astronomical Center, ul. Bartycka
  18, PL-00-716, Warszawa, Poland} 
\affiliation{Institute of Physics,
  Faculty of Philosophy and Science, Silesian University in Opava,
  Bezru{\v c}ovo n{\'a}m. 13, CZ-74601 Opava, Czech Republic}

\author{Jan Schee}\email[]{jan.schee@fpf.slu.cz}
\affiliation{Institute of Physics, Faculty of Philosophy and Science, Silesian
University in Opava, Bezru{\v c}ovo n{\'a}m. 13, CZ-74601 Opava,
Czech Republic}    

\author{W\l odek Klu\'zniak}\email[]{wlodek@camk.edu.pl}
\affiliation{Copernicus Astronomical Center, ul. Bartycka 18,                   %
               PL-00-716, Warszawa, Poland}  
\affiliation{Institute of Physics, Faculty of Philosophy and Science, Silesian
University in Opava, Bezru{\v c}ovo n{\'a}m. 13, CZ-74601 Opava, 
Czech Republic} 

\author{Zden\v{e}k Stuchl\'ik}\email[]{zdenek.stuchlik@fpf.slu.cz}
\affiliation{Institute of Physics, Faculty of Philosophy and Science, Silesian
University in Opava, Bezru{\v c}ovo n{\'a}m. 13, CZ-74601 Opava,
Czech Republic} 

\author{Marek Abramowicz}\email[]{marek.abramowicz@physics.gu.se}
\affiliation{Copernicus Astronomical Center, ul. Bartycka 18,
               PL-00-716, Warszawa, Poland} 
\affiliation{Institute of Physics, Faculty of Philosophy and Science, Silesian
University in Opava, Bezru{\v c}ovo n{\'a}m. 13, CZ-74601 Opava, 
Czech Republic} 
\affiliation{Physics Department, Gothenburg University,
               SE-412-96 G{\"o}teborg, Sweden}

\pacs{04.50.Kd, 04.70.Bw, 04.40.Dg, 95.10.Eg} 
                                                   %

           %
                    %


\begin{abstract}
We discuss photon and test-particle orbits in the Kehagias-Sfetsos
(KS) metric. For any value of the Ho{\v r}ava parameter $\omega$,
there are values of the gravitational mass $M$ for which the metric
describes a naked singularity, and this is always accompanied by a
vacuum ``antigravity sphere'' on whose surface a test particle can
remain at rest (in a zero angular momentum geodesic), and  inside
which no circular geodesics exist.  The observational appearance of an
accreting KS naked singularity in a binary system would be that of a
quasi-static spherical fluid shell surrounded by an accretion disk,
whose properties depend on the value of $M$, but are always very
different from accretion disks familiar from the Kerr-metric
solutions. The properties of the corresponding circular orbits are
qualitatively similar to those of the Reissner-Nordstr\"om naked
singularities.  When event horizons are present, the orbits outside
the  Kehagias-Sfetsos black hole are qualitatively similar  to those
of the Schwarzschild metric.
\end{abstract}

\maketitle

\section{Introduction}



Ho{\v r}ava \cite{horava2009} proposed a theory of gravity motivated
by a need to include quantum effects in the low-mass limit.  Its
action allows a spherically symmetric, asymptotically flat solution,
which has been found by Kehagias and Sfetsos \cite{KS} in  a modified
version of the theory that is compatible with
Minkowski vacuum. The solution involves a parameter, $\omega>0$, in
addition to the gravitational mass, $M$, and tends to the
Schwarzschild solution in the limit of large values of the dimensionless
parameter product $\omega M^2$.

Presumably, $\omega$ is a universal constant, if the solution found in
\cite{KS} correctly describes gravity.   However the properties of the
solution depend also on $M$.  For large masses, $\omega M^2>1/2$, the
Kehagias-Sfetsos solution has event horizons, i.e., it corresponds
 to a black hole \cite{KS}. 
 The current observational (lower) limits on $\omega$
 are not particularly stringent
 \cite{iorio2010, liu2011, harko2011tests, iorio2011},
and they are all compatible with the existence of stellar-mass
naked singularities.

 In this paper we discuss the circular
geodesics of the KS solution, basing our
work on the KS metric \cite{KS}, Eq.~(\ref{metric}), in which the
gravitating object is either a black hole or a naked singularity,
depending on its gravitational mass (once the Ho{\v r}ava parameter,
$\omega$, is fixed at any value).

Circular time-like geodesics are, of course, of great astronomical
interest, predating even the birth of physics.  Currently they play a
fundamental role not only in investigations of planetary motion, but
also in the dynamics of galaxies and of  the brightest X-ray sources,
such as quasars and X-ray binaries, where the emissions are powered by
accretion disks.  The properties of accretion disks, such as their
luminosity, photon spectrum, time variability and the possible
presence of an  inner edge of the disk close to the ISCO (innermost
stable circular orbit)  are all intimately related to
the properties of circular orbits of the accreting fluid 
 \cite{SS, NT, KMW, abramowicz05, olek}.

The astrophysical information is carried from the disk to the observer by photons travelling through the cosmic vacuum, so null 
geodesics are of great interest as well. Further, circular photon orbits, present already in the Schwarzschild solution to Einstein's 
general relativity (GR), may leave behind an observable time signature \cite{bursa, kazanas}.

Testing GR and alternative theories of gravity provides an aditional
motivation for the current study.  Geodesics around KS black holes
have already been studied in
 \cite{chen2010, abdujabbarov2011,  enolskii2011},
 as well as circular motion and  accretion disks in
slowly rotating extensions of the KS spacetime in \cite{harko2011}.
For any value of $\omega$, a sufficiently small $M$ value will allow
the naked singularity KS solution.
 Differences in the properties of accretion disks around black holes and putative naked singularites have been discussed in a different, GR, context \cite{harkovacs}.
 If the properties of the observed
stellar mass black hole candidates turn out to be incompatible with
the theoretical properties of the KS naked singularities, this will
provide a stringent lower limit to the value of the Ho{\v r}ava
parameter \cite{tests}.  Accordingly, we focus on the orbital
properties of the KS spacetime in the naked singularity case. Similar
investigations were recently  carried out for the Reissner-Nordstr\"om
metric \cite{stuchlik-02, pugliese2011}, and a remarkable qualitative
similarity between the properties of the circular geodesics of naked
singularities in GR and in Ho{\v r}ava's gravity is one of the
conclusions of our paper.

 \begin{figure}[h]
 \includegraphics[scale=0.65]{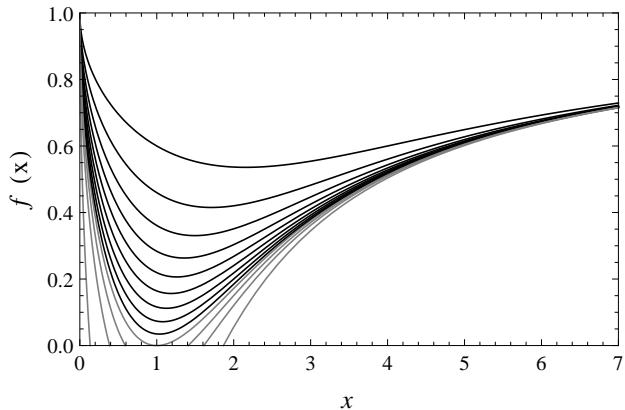} 
 \caption{\footnotesize The $g_{tt}$ coefficient of the KS metric.
For $L=0$, this is also the effective potential $V_{\rm eff}=f$,
 given by Eq. $(\ref{Veff-KS})$. 
 For the black curves (positive everywhere),
corresponding to naked singularities,
$\omega M^2$ varies from $0.05$
 to $0.45$ in steps of $0.05$, from the higher to the lower curve;
for the gray curves (having zeroes), corresponding to black holes,
 $\omega M^2$
has the values $ 0.5$, $ 0.6$, $0.8$, and $2.0$, increasing in the same sense.
 The function $f\le 1$ satisfies 
 $f\to 1$ as $x\to 0^+$
 and $x\to \infty$, and is continuous everywhere.
The zeroes of $f$ correspond to the radial location of the event horizons.}
 \label{fig:Vef,L=0}
\end{figure}    

\section{KS spacetime}

The Kehagias-Sfetsos (KS) solution corresponds to a spherically symmetric,
 static metric
    \begin{equation}
     ds^2 = g_{tt}dt^2  + g_{rr}dr^2  + g_{\theta\theta}d\theta^2
             + g_{\varphi\varphi}d\varphi^2,
    \end{equation}
and has the form \cite{KS}:
    \begin{equation}
     ds^2 = f(r)dt^2 - f^{-1}(r)dr^2 - r^2(d\theta^2 + \sin^2\theta d\varphi^2),
     \label{metric}
    \end{equation}
with
    \begin{equation}
     f(r) = 1 + r^2\omega\left[1-\Big(1+\frac{4M}{\omega r^3}\Big)^{1/2}\right].
    \label{fKS}
    \end{equation}   
The Schwarzschild solution $f(r) = 1 -2M/r$ 
is recovered in the limit of $\omega\rightarrow\infty$.
Further insight into the properties of the metric may be gleaned
from embedding diagrams \cite{embed}.

At first sight there are two parameters in the metric coefficients, $\omega$ and $M$. In fact, the KS metric depends on a single parameter alone, 
the combination $\omega M^2$. Indeed, in
terms of $x\equiv r/M$,
   \begin{equation}
    f(x) = 1 -
      \omega M^2 x^2\left[\Big(1+\frac{4}{\omega M^2 x^3}\Big)^{1/2}-1\right].
   \label{fxKS}
   \end{equation}
For convenience, we extend the domain to include $x=0$, where we take
$f(0)=1$. Note that $f\le 1$ and is continuous for all $x\ge0$.
 The behavior of $f$ can be seen in
Fig. \ref{fig:Vef,L=0}, for different values of $\omega M^2$.
As $f\equiv g_{tt}$ here, $f=0$ corresponds to the presence of an event
horizon.  This occurs \cite{KS} for $\omega M^2\ge 1/2$, with two horizons
present for all $\omega M^2> 1/2$. 

 For $\omega M^2<1/2$ the
KS solution has an interesting feature, which distinguishes it
from naked singularity solutions in GR:  $g_{tt}$ is finite
everywhere. The metric coefficients are continuous at $r=0$, with
$f(0)=1$. However, their radial derivatives are not finite:
$f'(x)\to\infty$ as $x\to 0$, and this leads to a singularity in the
curvature \cite{KS}.

The radial acceleration of a static time-like observer is given by  
  \begin{equation}
   a_r=\frac{1}{2f} f'(r).
  \end{equation}
The general properties of the KS metric imply that this acceleration
vanishes at a certain radius in the naked singularity case ($\omega M^2<1/2$).
 Indeed, since $f$ is continuous and
$f(0)=1$, $f(\infty)=1$, it follows from the Rolle-Lagrange theorem
that there will be a radius at which $f'=0$.
Further, as  $f\le1$, this radius corresponds to a minimum of $f$
(Fig.~\ref{fig:Vef,L=0}), and this is a stable equilibrium position of
any test particle.
This radius,
$r_{\rm 0G}$, at which $a_r=0$, could be termed the ``antigravity radius''.
Its value is 
  \begin{equation}
  x_{\rm 0G}\equiv r_{\rm 0G}/M=(2\omega M^2)^{-1/3}.
  \label{og}
  \end{equation}
The angular momentum and frequency of $r=\,{\rm const}$
geodesics also vanish at $r_{\rm 0G}$. Below this radius the
radial  acceleration of a static observer points inwards and
therefore there are no circular time-like geodesics, this is the case
of  repulsive gravity (antigravity).  Thus, circular geodesics exist
only for $r>r_{\rm 0G}$, and there are also stable time-like geodesics
corresponding to rest at  $r=r_{\rm 0G}$.

\section{Equatorial geodesic motion}
To fix the notation we begin with a brief review of circular motion in a spherically symmetric, static metric, with signature (+  -- -- -- ).
For geodesics we have the following constants of motion:
    \begin{equation}\label{energy}
     E\equiv g_{tt}\dot{t} = u_t,
    \end{equation}

    \begin{equation}\label{angularmomentum}
     L\equiv -g_{\varphi\varphi}\dot{\varphi} = -u_\varphi,
    \end{equation}
respectively, the energy and angular momentum of a test particle. 
From these conserved quantities we can construct the
 {\it specific} angular momentum, also conserved,


    \begin{equation}\label{specificangularmomentum}
     \ell\equiv \frac{L}{E} = -\frac{u_\varphi}{u_t}.
    \end{equation}   

For circular equatorial geodesics ($r=\,$const, $\theta=0$),
 we may write the specific angular momentum $\ell^2(r)$ 
in terms of its angular velocity $\Omega$:
  \begin{equation}\label{ellOmega}
   \ell^2(r) = \tilde{r}^4\Omega^2(r),
  \end{equation}
where (e.g., \cite{abramowicz05})
  \begin{equation}
   \tilde{r}^2=-\frac{g_{\varphi\varphi}}{g_{tt}}
   \label{rtilde}
  \end{equation}
and 
  \begin{equation}
   \Omega^2(r)=-\frac{\partial_r g_{tt}}{\partial_r g_{\varphi\varphi}}.
  \end{equation}    
\begin{figure}[h]
 \includegraphics[scale=0.65]{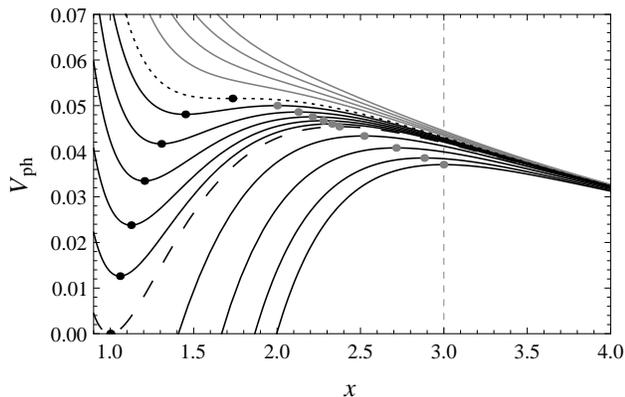} 
 \caption{\footnotesize Effective potential for photons at the equatorial
 plane, Eq. (\ref{Vph_f}). 
 Different curves correspond to different values of $\omega M^2$.
 The upper, gray, curves have $\omega M^2=$  0.30, 0.32, 0.34, 0,36,
 from top to bottom.  They do not allow circular photon orbits.
 The black short-dashed curve corresponds to $(\omega M^2)_{\rm ms.ph}$,
 Eq. (\ref{omegacritphoton}). 
 The positive black curves have $\omega M^2$ varying from 0.40 to 0.48
 in steps of 0.02 and correspond to the naked singularity case.
  The dashed black curve corresponds to $\omega M^2=0.5$.
 The black curves below it have  $\omega M^2=$ 0.6, 0.9, 2.0 and 100.
 Black dots correspond to stable photon orbits,
 whereas gray dots correspond to unstable photon orbits.
 The black dot on the short-dashed curve represents
 the marginally stable photon orbit.
 The vertical dashed line at $x=3$ marks the Schwarzschild
  value of the photon orbit radius.
 }
 \label{fig:Vph}
\end{figure}      

\subsection{Time-like circular geodesics}
    
The radial dependence of time-like equatorial geodesic motion
 is described by the equation
  \begin{equation}
 E^2 = -g_{tt}g_{rr}\dot{r}^2 + V_{\rm eff}(r),
  \end{equation}
where the {\it effective potential} is defined by
  \begin{equation}
   V_{\rm eff}=\Big(1-\frac{L^2}{g_{\varphi\varphi}}\Big)g_{tt}. 
  \end{equation}

From $\partial V_{\rm eff}/\partial r=0$ we obtain the angular momentum of circular geodesics:
  \begin{equation}\label{L2circ-axi}
   L^2(r)=-\frac{g_{\varphi\varphi}^2 g_{tt,r}}{g_{tt}g_{\varphi\varphi,r}-g_{\varphi\varphi}g_{tt,r}}
  \end{equation}
This formula was obtained in \cite{letelier03} by a direct analysis of the geodesic equations.

The second derivative of the effective potential is related to the
stability of circular motion. Indeed, $\partial^2 V_{\rm eff}/\partial r^2$
  evaluated at the circular orbit is proportional to the radial
epicyclic frequency and hence its sign determines whether or not the
Rayleigh criterion is satisfied \cite{abramowicz05}.  This can also be
seen from the relation

  \begin{equation}
   \frac{\partial^2 V_{\rm eff}}{\partial r^2}\bigg|_{\rm circ}
        =\frac{g_{tt,r}}{L^2(r)}\frac{dL^2(r)}{dr},
  \end{equation}
valid in the equatorial plane circular orbits of any static, 
axially symmetric spacetime.
For time-like circular geodesics, 
$E(r)$ and $L(r)$ satisfy a precise relation \cite{letelier03}
 $dL^2/dr = \tilde r^2 dE^2/dr$, and hence
 $\ell^2(r)$ and $L^2(r)$ satisfy 
  \begin{equation}
   \frac{d\ell^2(r)}{dr}
   =\frac{1}{E^2}\Big(1-\big[\tilde{r}\Omega(r)\big]^2\Big)\frac{dL^2(r)}{dr}.
   \label{stability}
  \end{equation}
Thus,  for $r>r_{\rm 0G}$
the sign of $\partial^2 V_{\rm eff}/\partial r^2$  is the same as that of
$d\ell^2(r)/dr$ and of $dL^2(r)/dr$. As is well known,
radial stability corresponds to $d\ell^2(r)/dr>0$, or equally to
$dL^2(r)/dr>0$, while
radial instability corresponds to $d\ell^2(r)/dr<0$, i.e. to $dL^2(r)/dr<0$.
The term between brackets in Eq.~(\ref{stability})
 is recognized as the particle's circular velocity
measured by a local static observer,
$v^2=\tilde{r}^2\Omega^2$.
 
\subsection{Null circular geodesics}

The motion   of photons in the equatorial plane is described by
the  equation
  \begin{equation}
 \frac{E^2}{L^2} = -g_{tt}g_{rr} \frac{\dot{r}^2}{L^2} + V_{\rm ph}(r),
  \end{equation}

with the effective potential
  \begin{equation}\label{Vph-axi}
   V_{\rm ph}=-\frac{g_{tt}}{g_{\varphi\varphi}} = \frac{1}{\tilde{r}^2}.
  \end{equation}
The condition for a circular photon orbit is $\partial V_{\rm ph}/\partial r=0$.
Maxima of $V_{\rm ph}$ correspond to radially unstable photon orbits, whereas minima of $V_{\rm ph}$ correspond to stable photon orbits.
From (\ref{rtilde}),  (\ref{Vph-axi}), and (\ref{L2circ-axi}) we see that in any static, axially symmetric spacetime 
$L^2(r)\to\infty$ as the radius of the circular time-like geodesic
approaches that of the photon orbit.

\begin{figure}[h]
 \includegraphics[scale=0.65]{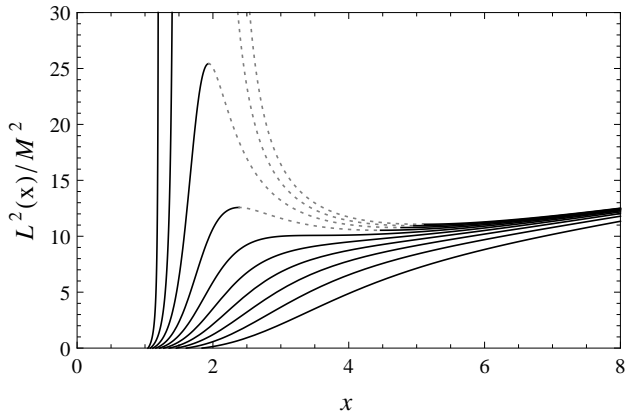} 
 \caption{\footnotesize Angular momentum of test particles in circular
   orbits as a function of $x$. Different curves are for different
   values of $\omega M^2$, varying from 0.08 (bottom) to 0.44 (top) in
   steps of 0.04. Solid black lines correspond to stable motion,
   whereas dotted gray lines correspond to  unstable motion.  The
   lower six curves correspond to the case of no marginally stable
   orbits, $\omega M^2<(\omega M^2)_{\rm ms}$. The next two (middle)
   curves, for
    $(\omega M^2)_{\rm ms}<\omega M^2<(\omega M^2)_{\rm ms.ph}$,
    exhibit two marginally stable orbits,
    an ISCO and an OSCO (outermost stable circular orbit).
  For the upper two curves, $(\omega M^2)_{\rm ms.ph}< \omega M^2$, the
  angular momentum in the inner stable branch goes to infinity at the
  stable photon orbit, and the angular momentum in the unstable
  branch goes to infinity at the unstable photon orbit.  There are no
  circular orbits between the photon orbits. See \S \ref{Geodesic} and
  Fig.~\ref{fig:radii_w}.
  }
 \label{fig:ell2-x}
\end{figure}  

\section{Geodesic motion in KS spacetime}
\label{Geodesic}

In KS spacetime, with the metric given by eqs. (\ref{metric}--\ref{fKS}), we have
   
  \begin{equation}\label{Vph_f}
   V_{\rm ph}=\frac{f}{r^2}
  \end{equation}

 \begin{equation}\label{Veff-KS}
   V_{\rm eff}=f(r)\Big(1+\frac{L^2}{r^2}\Big),
  \end{equation}   
  
  \begin{equation}
   L^2(r)=\frac{r^3f'(r)}{2f-rf'},
  \end{equation}
  
  \begin{equation}\label{ell2rf}
   \ell^2(r) = \Big(\frac{r^3}{2f^2}\Big) f'(r),
  \end{equation}  
  
  \begin{equation}\label{omega2f}
   \Omega^2(r)=\frac{f'(r)}{2r}.
  \end{equation}

\begin{figure}[h]
 \includegraphics[scale=0.65]{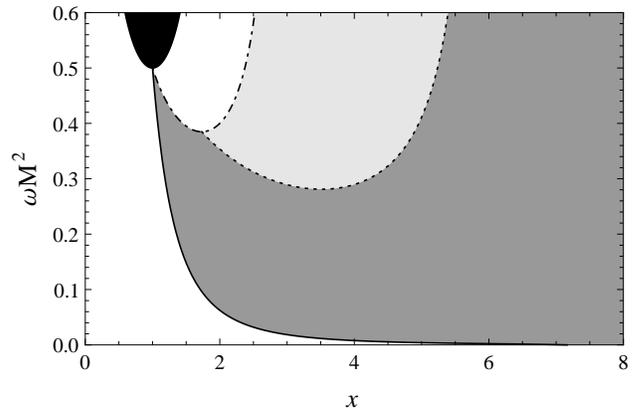} 
 \caption{\footnotesize Regions of (in)stability of time-like circular
  geodesics and relevant radii in the KS spacetime.  Black:
   region between horizons, where the versor $\partial_t$ is
     space-like.  Dark gray: stability region for time-like circular
   geodesics.  Light gray: instability region for time-like circular
   geodesics.  White: region of no circular time-like geodesics. Solid
   line: $x_{\rm 0G}$. Dotted lines: $x_{\rm ms}$.  Dot-dashed lines:
   $x_{\rm ph}$.  We see that the $x_{\rm ms}$ line meets the
   photon-orbit lines, at ($x_{\rm ms.ph},(\omega M^2)_{\rm ms.ph}$),
   see Eq. (\ref{omegacritphoton}).  }
 \label{fig:radii_w}
\end{figure}  

The allowed regions for circular time-like motion have qualitatively
different properties, depending on the value of $\omega M^2$, as
summarized in Fig. \ref{fig:radii_w}.  This figure shows the
antigravity radius $x_{\rm 0G}$ (solid line), the marginally stable orbits
$x_{\rm ms}$ (dotted lines) and the photon orbits $x_{\rm ph}$
(dot-dashed lines). We also show the different regions of stability
for circular time-like geodesics: stable region (dark gray), unstable
region (light gray) and ``forbidden'' regions where no circular
geodesics are allowed (white).  There are two such forbidden regions:
one between the two photon orbits (at lower values of $\omega M^2$) or
between the outer horizon and the unstable photon orbit (at higher values of
$\omega M^2$), and one where static observers suffer an inward radial
acceleration (either the region  $x<x_{\rm 0G}$,  at lower values of
$\omega M^2$, or the one between the singularity and the inner
horizon, at higher values of $\omega M^2$).  The behavior of
$L^2(r)$ is depicted in Fig.~\ref{fig:ell2-x}; note the correspondence
of the curves with the stability regions  presented in
Fig.~\ref{fig:radii_w}.   Since the spacetime is spherically symmetric,
all circular time-like geodesics are vertically stable.
 
The photon radii in KS spacetime are  positive solutions of the equation
  \begin{equation}\label{photoneq}
   x_{\rm ph}^3 - 9x_{\rm ph} +\frac{4}{\omega M^2} =  0,
  \end{equation}  
and the two photon orbits merge at     
  \begin{equation}\label{omegacritphoton}
   (\omega M^2)_{\rm ms.ph}=\frac{2}{3\sqrt{3}}=0.384900,
  \end{equation}
for $x_{\rm ms.ph}=\sqrt{3}$.
An interesting feature of photon orbits in the KS naked singularity
spacetime is that the inner photon  orbit is stable. This fact can be
seen directly from the behavior of $V_{\rm ph}$, Eq. (\ref{Vph_f}),
and it is illustrated in Fig. \ref{fig:Vph}. Related to this fact is a
potentially relevant astrophysical consequence:  for values of $\omega
M^2$ which admit the stable photon orbit there is also an inner region
of stable time-like circular geodesics, as depicted in
Fig. \ref{fig:radii_w}. This region of stability (with respect to both
radial and vertical perturbations)  extends from $x_{\rm 0G}$ to the inner
photon radius.

The point $(\sqrt{3},2\sqrt{3}/9)$ on the $x$-$\omega M^2$ plane 
at which the stable and unstable circular photon orbits merge
into the marginally stable photon orbit, Eq. (\ref{omegacritphoton}),
is also the termination point
of the locus of the marginally stable time-like orbits,
as can be seen in Fig. \ref{fig:radii_w}.
For $\omega M^2<(\omega M^2)_{\rm ms.ph}$ there are no photon orbits.

For $(\omega M^2)_{\rm ms}<\omega M^2<(\omega M^2)_{\rm ms.ph}$, with
 $(\omega M^2)_{\rm ms}=0.281100$,  there are two marginally stable orbits
among the time-like circular geodesics.
 The inner one, at radius 
$r_{\rm OSCO}$ could be termed the 
OSCO as for all $r$ satisfying
$r_{\rm 0G}<r<r_{\rm OSCO}$ there are stable circular orbits,
while the outer one, at $r_{\rm ISCO}$,
 present for all $\omega M^2>(\omega M^2)_{\rm ms}$ is
the ISCO familiar from the Schwarzschild metric, as there are stable
circular orbits for all $r>r_{\rm ISCO}$.
Thus, there are two regions of stability
of circular test-particle motion for
 $(\omega M^2)_{\rm ms}<\omega M^2<(\omega M^2)_{\rm ms.ph}$.
The OSCO and ISCO coincide for $\omega M^2=(\omega M^2)_{\rm ms}$.
For $\omega M^2<(\omega M^2)_{\rm ms}$ all circular orbits are stable down to
$x_{\rm 0G}$.

This situation is analogous to the inner stability regions found in
Reissner-Nordstr\"om naked-singularity spacetimes \cite{pugliese2011}
and Reissner-Nordstr\"om spacetimes with a cosmological constant
\cite{stuchlik-02}, which may suggest that naked singularities are
somehow related to this inner region of stability.

\section{The effective potential for test particles}

In this section, we systematically discuss the behavior of
 $V_{\rm  eff}$ for test-particle motion in terms of  $\omega M^2$.

\begin{figure}[h]
 \includegraphics[scale=0.65]{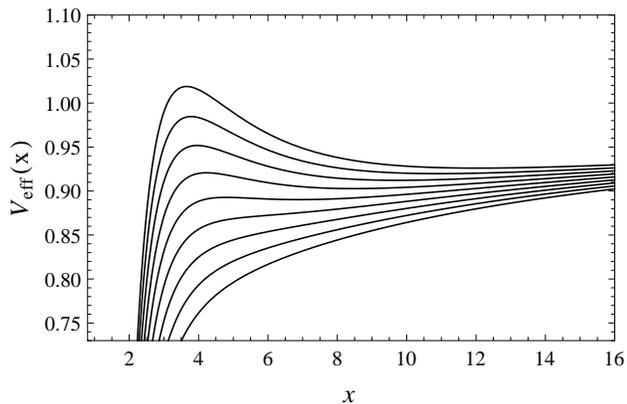} \quad
 \caption{\footnotesize Effective potential $V_{\rm eff}$ as a function
 of $x$ for different values of $L^2$ (decreasing from top to bottom).
Depicted in the figure is the KS black-hole case, $\omega M^2=1>1/2$,
 qualitatively similar to the Schwarzschild solution.
 }
\label{fig:Veff1}
\end{figure}  

For the black-hole case ($\omega M^2>1/2$)
 the situation is analogous to the Schwarzschild metric
(Fig. \ref{fig:Veff1}). For sufficiently high values of angular
momentum $V_{\rm eff}(r)$ has a maximum at $r_{\rm unst}$,
corresponding to an unstable circular orbit, and a minimum at $r_{\rm st}$,
corresponding to a stable circular orbit. At a certain critical value of
angular momentum, $L_{\rm ms}$, the two extrema merge at the radius
$r_{\rm ms}$ of the marginally stable circular orbit. The unstable, marginally
stable, and stable orbit radii satisfy 
$r_{\rm unst}<r_{\rm ms}<r_{\rm  st}$. Hence, in this case the marginally stable orbit
is an ISCO.
 For $L<L_{\rm ms}$ no circular orbits are possible.

\begin{figure}[h]
 \includegraphics[scale=0.65]{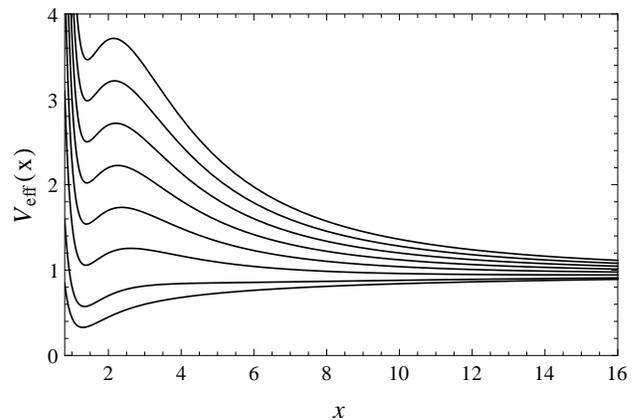} \quad
 \caption{\footnotesize Same as Fig.~\ref{fig:Veff1}, but
for $(\omega M^2)_{\rm ms.ph}<\omega M^2=0.4<0.5$, a regime with two
photon orbits.
 }
\label{fig:Veff2}
\end{figure}  

\begin{figure}[h]
 \includegraphics[scale=0.65]{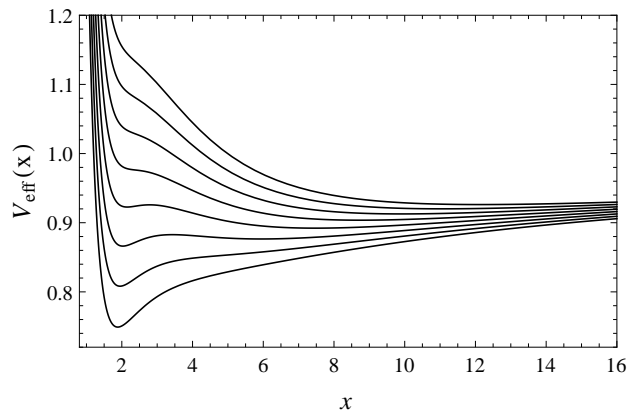}
\caption{\footnotesize Same as Fig.~\ref{fig:Veff1}, but
for $(\omega M^2)_{\rm ms}<\omega M^2=0.32<(\omega M^2)_{\rm ms.ph}$, the regime with two marginally stable orbits.
 }
\label{fig:Veff3}
\end{figure}

\begin{figure}[h]
 \includegraphics[scale=0.65]{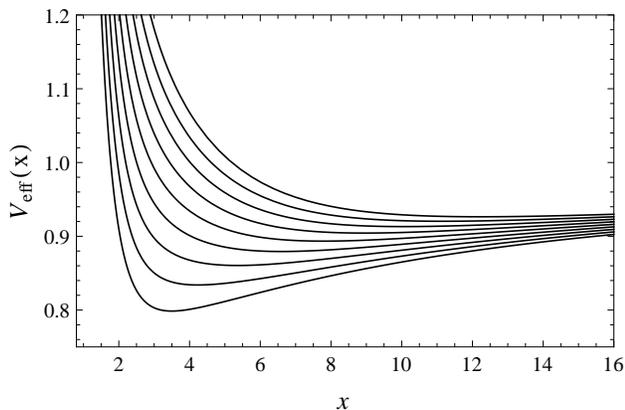}
\caption{\footnotesize Same as Fig.~\ref{fig:Veff1}, but
for $\omega M^2=0.2<(\omega M^2)_{\rm ms}$, the regime of stable orbits alone,
 qualitatively similar to Newtonian gravity for $x>x_{\rm 0G}$, c.f. Eq.~\ref{og}.
 }
\label{fig:Veff4}
\end{figure}

The remaining three cases
 (Figs.~\ref{fig:Veff2}, \ref{fig:Veff3}, \ref{fig:Veff4}) 
correspond to naked singularities. 

If $(\omega M^2)_{\rm ms.ph}<\omega M^2<1/2$ the spacetime admits two
photon orbits, which correspond to the $L^2\to\infty$ limit of test-particle orbits  (see Figs.~\ref{fig:Vph}-\ref{fig:radii_w}).  This is indeed the
situation for the first two extrema of $V_{\rm eff}$
(Fig. \ref{fig:Veff2}). As $L^2$ increases (without bounds), both the
minimum and maximum values of $V_{\rm eff}(r)$ increase (without
bounds). The marginally stable orbit appears (at a radius larger than
that of the unstable photon orbit) in the minimum of $L^2(r)$, at
$L^2_{\rm ms}(\omega M^2)$.  Note that a stable circular orbit is
always present between $x_{\rm 0G}$ and the stable photon orbit, while the
unstable circular orbit and the second (familiar) stable circular
orbit (both at radii larger than that of the unstable photon orbit)
only for values of $L^2>L^2_{\rm ms}$.
  
We note that the behavior of $V_{\rm eff}$ for high angular momentum,
as shown in  Fig.~\ref{fig:Veff2}, is not a  particular feature of the
KS naked-singularity solution, but instead a property of geodesic
motion. If there is a local  maximum of $V_{\rm eff}$ at
 $r_{\rm  unst}(L^2)$ for every high value of $L^2$ and if this sequence of
radii has a bounded limit as $L^2\rightarrow\infty$,
 this limit will  correspond to an unstable
circular photon orbit. In the same way, if there is a local  minimum
of $V_{\rm eff}$ at $r_{\rm unst}(L^2)$ for every high value of $L^2$
and if this sequence of radii has a bounded limit
 as $L^2\rightarrow\infty$, this limit will
correspond to a stable circular photon orbit. 

If we now consider $(\omega M^2)_{\rm ms}<\omega M^2<(\omega M^2)_{\rm
  ms.ph}$ there still are two separate regions  of stability, but no
photon orbit (Fig.~\ref{fig:radii_w}). 
 We see from Fig.~\ref{fig:Veff3} that the inner stable
orbit is present for low values of $L^2$, but as the magnitude of angular
momentum is increased a marginally stable orbit is reached
 (the OSCO, see Fig.~\ref{fig:ell2-x}),
and at higher values still of $|L|$ only the familiar outer stable
orbits persist.  Conversely, the outer stable orbits exist only for
values of $|L|$ larger than the value in the ISCO.  The unstable
circular orbits exist only in  a range of $L^2$ values, merging with
the outer stable circular orbit at the lower end of the range
(corresponding to the ISCO) and with the inner circular orbit at the
higher end of the range (corresponding to the OSCO).

\begin{figure}[h]
 \includegraphics[scale=0.65]{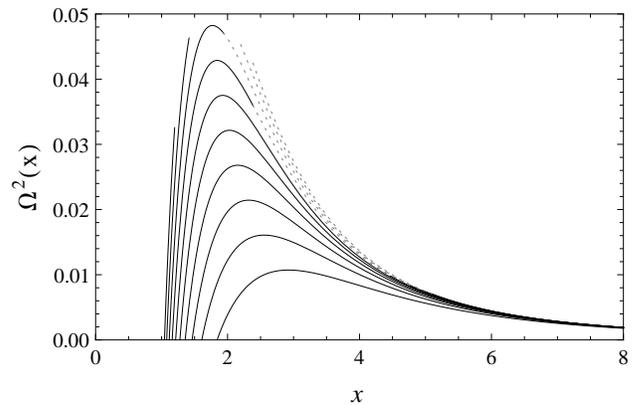} 
 \caption{\footnotesize Angular frequency of the circular orbits as a function of $x$. Different curves are for different values of $\omega M^2$
 in the naked-singularity regime,
 varying from 0.08 (bottom) to 0.44 (top) in steps of 0.04. Solid black lines correspond to stable motion, whereas dotted gray lines correspond to 
 unstable motion.
 }
 \label{fig:omega2-x}
\end{figure}

The last case is $\omega M^2<(\omega M^2)_{\rm ms}$, for which the
situation is analogous to ubiquitous
motion in Newtonian gravity at $x>x_{\rm 0G}$.
Fig.~\ref{fig:Veff4} shows that in this regime there is only one circular
orbit for each value of angular momentum, and this orbit is stable.
The orbits extend down to  $x_{\rm 0G}$,
 which plays the role of the radius of a Newtonian
star, or planet.
However, it would be misleading to expect that the properties of the orbits
as a function of their radius are the same as that of Keplerian orbits
in a Newtonian $1/r$ potential, where the orbital frequency monotonically
increases with $r$. This cannot be the case here, as the angular
frequency of the orbits goes to zero at both edges of the allowed
domain, $\Omega\rightarrow0$
 equally for $r\rightarrow\infty$ and  $r\rightarrow r_{\rm 0G}$, implying
that the (positive) angular frequency (squared) has a maximum. 
This is indeed the case,
as can be seen in  Fig.~\ref{fig:omega2-x}, which shows $\Omega^2(r)$.

The profiles of the angular frequency $\Omega(x)$, given by Eq. (\ref{omega2f}),
in the three naked-singularity regimes defined by the values of $\omega M^2<1/2$
are all similar. 
$\Omega^2(x)$ is always a curve with a zero at the gravity radius
 $x_{\rm 0G}$, it then monotonically increases, passes through a maximum---except
 for  $(\omega M^2)_{\rm ms.ph}<\omega M^2<1/2$, when there is
a region without circular orbits, so the domain of $\Omega^2$ 
is not simply connected---and
 decreases monotonically tending
 to zero at infinity, $\Omega(x)\rightarrow0$ for $r\rightarrow\infty$.
For $(\omega M^2)_{\rm ms}<\omega M^2<(\omega M^2)_{\rm ms.ph}$, a part of the curve
corresponds to unstable orbits.

\section{Discussion}
The KS naked singularity in Ho{\v r}ava's gravity
 has properties which distinguish it from
those of the familiar Schwarzschild and Kerr solutions of GR.
First, since $g_{tt}$ at the singularity has the same value as at infinity,
the singularity can be reached by a zero angular momentum test particle
dropped from rest at
infinity. However, if the particle loses some energy on its
inbound trajectory, it will be reflected back and will oscillate between
two turning points. If the energy loss continues, the particle
 will eventually settle at the minimum of the effective potential
for radial motion of a test particle (Fig.~\ref{fig:Vef,L=0}),
i.e., at the minimum of $g_{tt}$.

By the same token,  since $g_{tt}$ is finite in the neighbourhood of the
singularity, $f$ being continuous, with $f(0)=1$ (Eq.~[\ref{fxKS}]),
 the effective potential for nonradial motion ($L\neq0$)
 will always present  an infinite centrifugal barrier, so a particle carrying 
any angular momentum will never reach the singularity unless it manages to lose
its angular momentum without losing energy. A more likely outcome is that the
particle will lose both angular momentum and kinetic energy,
in which case it also can
settle at the minimum of $g_{tt}$. In view of spherical symmetry of the KS
metric, this minimum has the form of the surface of a sphere. Inside this
``antigravity sphere'' static observers suffer an inward radial acceleration
(an outward force).

We thus come to a most interesting conclusion. Accretion of matter by the naked
KS singularity will result in accumulation of matter on the surface of
a vacuum sphere around the naked singularity. Thus, an astrophysically relevant
KS naked singularity will present the aspect of a spherical surface
at the top of a shell of
accumulated matter. For stellar-mass objects, the sphere will have
a radius close to that of the gravitational radius of the naked singularity,
$r\sim M$.

If the naked singularity is in a binary system, it may accrete through a disk.
It is thought that viscous dissipation in accretion disks is a result
of MRI (the magnetorotational instability) \cite{mri}, which occurs as long
as angular frequency increases outwards, $d\Omega^2/dr>0$.
As we can see in Fig.~\ref{fig:omega2-x}, this would occur only down to the
maximum of $\Omega^2$.
The maximum of $\Omega^2(x)$ is rather close to the naked singularity.
Indeed, for $\omega M^2$ within an order of magnitude or two of
 the black hole limit ($\omega M^2=1/2$), the maximum is at $M<r<6M$.

For $\omega M^2<(\omega M^2)_{\rm ms}$ the orbits are stable down to the
antigravity radius
 (Figs.~\ref{fig:radii_w}, \ref{fig:Veff4}, \ref{fig:omega2-x}),
 so in this case we would have a quasistatic shell
of fluid (accreted matter) on the surface of the hollow (vacuum) antigravity
sphere concentric with the naked singularity, a cold, possibly opaque disk,
abutting the sphere,
and a hot accretion disk starting at a few gravitational radii.
From the astronomical point of view, this system would present an
aspect quite different from a Schwarzschild or Kerr black hole with an accretion
disk with a hole in it (close to the ISCO) through which a second order lensed image of the disk can be observed. This suggests that a stringent constraint can be placed on the  Ho{\v r}ava parameter through observation of astrophysical
black holes (\cite{tests}).

\section{Conclusions}
We discussed photon and test-particle orbits in the Kehagias-Sfetsos
metric as a function of the Ho{\v r}ava parameter times mass squared,
$\omega M^2$. For naked singularities there are three
different regimes of circular motion, depending on the value of $M$.
There is always an ``antigravity sphere'' on whose surface a test
particle can remain at rest (in a zero angular momentum geodesic), and
inside which no circular geodesics exist.  For low values of the mass,
circular motion is possible everywhere outside the antigravity sphere
and is always stable. For intermediate mass values, there are two distinct
regions of stable circular motion---the inner one terminates in an
outer marginally stable circular orbit (OSCO) and the outer one begins
in an inner  marginally stable circular orbit (ISCO)---the two regions
are separated by a region where circular orbits are unstable.  For the
largest values of $M$  compatible with the existence of a naked
singularity there is an inner region of  stable time-like circular
geodesics, terminated by a stable photon orbit, and an outer region of
stable circular geodesics beginning with an ISCO.  Between the stable
photon orbit and the ISCO there are two regions separated by an
unstable circular photon orbit: in the inner one no circular orbits
are allowed, and in the outer one circular time-like geodesics are
unstable. The KS metric is static, and this is the only one we have considered.
In GR it is known that the properties of orbits for rotating naked
singularities differ from those of the static ones \cite{1980}.

 The properties of circular orbits of the  Kehagias-Sfetsos
and the Reissner-Nordstr\"om naked singularities are  qualitatively
similar (c.f., this work and \cite{stuchlik-02, pugliese2011}),
but very different from the black hole ones.
  When event horizons are present, the orbits outside the
black hole are qualitatively similar to those of the Schwarzschild
metric.  No circular geodesics exist between the singularity and the
inner horizon.
The differences between the KS naked singularity
and GR black hole cases may have astrophysical consequences,
which we intend to explore in forthcoming papers.
In particular, they may allow placing more stringent limits on the
Ho{\v r}ava parameter (\cite{tests}).

\acknowledgments
This work supported in part by Brazilian grant from
 ``Funda\c{c}\~ao de Amparo \`a Pesquisa do Estado de S\~ao Paulo (FAPESP)'',
 no. 2013/01001-0, Czech grantCZ.1.07/2.3.00/20.0071 ``Synergy''
and Polish NCN grants UMO-2013/08/A/ST9/00795, UMO-2011/01/B/ST9/05439.

\end{document}